\pdfoutput=1
\pdfoutput=1
\pdfoutput=1
\pdfoutput=1
\pdfoutput=1
\documentclass[preprint,showkeys,unsortedaddress,superscriptaddress, amsfonts,showpacs,amsmath,amssymb]{revtex4}
\usepackage[dvips]{color}
\usepackage{epsfig}
\usepackage{amsmath}
\usepackage{graphicx}

\begin{document}
\title{Tachyonic Intermediate Inflation in DGP Cosmology with new observations}

\author{A. Ravanpak}
\email{a.ravanpak@vru.ac.ir}
\affiliation{Department of Physics, Vali-e-Asr University, Rafsanjan, Iran}
\author{H. Farajollahi}
\email{hosseinf@guilan.ac.ir}
\affiliation{Department of Physics, University of Guilan, Rasht, Iran}
\affiliation{School of Physics, University of New South Wales, Sydney, NSW, 2052, Australia}
\author{G. F. Fadakar}
\email{gfadakar@guilan.ac.ir}
\affiliation{Department of Physics, University of Guilan, Rasht, Iran}

\date{\small {\today}}

\begin{abstract}

In this article we study an intermediate inflationary model in the context of Dvali-Gabadadze-Porrati (DGP) cosmology caused by a tachyon scalar field. Considering slow-roll inflation we discuss the dynamics of the Universe. Using perturbation theory, we estimate some of the model parameters numerically and compare them with the recent results from Planck satellite.

\end{abstract}


\maketitle

\section{Introduction}

Inflation is a very short epoch at very early stages of the history of the Universe in which the Universe experiences a very rapidly accelerated expansion. This phase first proposed in \cite{Guth}, to solve some of the problems of the hot big bang model of cosmology, such as flatness problem, horizon problem and monopole problem. This model of inflation improved gradually to a more accurate scenario called slow-roll inflation which supports a long enough period of inflation \cite{Linde},\cite{Albrecht}. But maybe, the true merit of inflation is that it provides some inhomogeneities in the Universe arisen from vacuum fluctuations and so can explain the large scale structure of the Universe and the anisotropies in the cosmic microwave background (CMB) radiation \cite{Liddle},\cite{Dodelson}.

There are several inflationary models which differ in their expression of scale factor parameter, $a(t)$. Among them, the intermediate inflation is of particular interest because it arises from an effective theory at low dimension of a more fundamental string theory \cite{Sanyal}. In this class of inflationary models, the scale factor varies with time faster than the scale factor of power law inflation in which $a(t)=t^p; p>1$, but still slower than the scale factor of standard de-Sitter inflation in which $a(t)=\exp(Ht)$, where $H=\dot a/a$, is the Hubble parameter and dot means derivative with respect to cosmic time $t$. The behavior of scale factor parameter in terms of time in intermediate inflationary models is expressed as
\begin{equation}\label{scalefactor}
    a(t)=\exp(At^f),
\end{equation}
in which $0<f<1$ and $A>0$, are constant parameters.

In general, inflation drives by the potential of a standard scalar field, the inflaton, where it obeys the Klein-Gordon (KG) equation. But, there is a non-standard scalar field action motivated from string theory which can be used to drive an inflationary phase, called tachyon field \cite{Sen1}-\cite{Steer}. Because its equation of state parameter is bounded as, $-1<w<0$, it can play the role of inflaton field, well \cite{Mazumdar}-\cite{Sami2}. Tachyon field has some other applications in cosmology, too. It can play the role of dark sectors of the Universe \cite{Padmanabhan}-\cite{Ravanpak}. Also, in \cite{Gibbons}, it has been shown that the tachyon field can drive inflation and then behave as dark matter or a non-relativistic fluid. In principle, the tachyon inflation is a $k$-inflationary model \cite{Garriga}, with its own features. It has a positive potential $V(\phi)$, where has a maximum at $\phi = 0$ and approaches zero when $|\phi|\rightarrow \infty$. Meanwhile, during the entire of this process, $\frac{dV(\phi)}{d\phi}<0$.

It might be useful to deals with the end of inflation and a mechanism called reheating in which the temperature of the Universe grows in many orders of magnitude to recover the Big-Bang cosmology. Conventionally, this happens when inflaton field starts to oscillate around the minimum of its potential. At this stage, most of the matter and radiation of our Universe is created via the decay of the inflaton field. But reheating in all inflationary models inspired by string theory is problematic \cite{Kofman}. For instance, one of the characteristics of intermediate inflationary models is that there is not such a minimum in their inflaton potential \cite{Campo1m}. Also, as we mentioned earlier the potential of the tachyon field does not present a minimum at a finite time, too. Thus, in such a models we need a different mechanism to bring inflation to an end. There are a few other reheating mechanisms \cite{Cline}-\cite{Felder}, among them the curvaton scenario \cite{Lyth}, has been more successful to solve the problem in tachyonic and intermediate inflationary models \cite{Campuzano}-\cite{Farajollahi10}.

On the other hand, the theory of extra dimensions which has come out of the string theory, has attracted a great amount of attention in the past two decades. Several five dimensional (5D) cosmological models have been proposed to explain the weakness of gravity and hierarchy problem \cite{Arkani}-\cite{Randall2}. In these models our four dimensional (4D) Universe is a surface dubbed brane, embedded into a higher dimensional bulk spacetime. It is assumed that the standard model of particle physics is confined to the brane and only gravitons can propagate into the bulk. The important effect of considering extra dimensions is that they modify the Friedmann equations by adding some new terms. These theories, specially the Randall-Sundrum (RS) type II model, with a fine-tuning relation as $\Lambda_4=\frac{1}{2}(\Lambda_5+\frac{1}{6}\kappa_5^4\lambda^2)$, in which $\Lambda_4$, $\Lambda_5$, $\kappa_5^2$ and $\lambda$, are cosmological constant on the brane, cosmological constant of the bulk, five dimensional gravitational constant and the tension of the brane respectively, have been widely utilized in the literature to explain the dynamics of the Universe, more precisely \cite{Sami2},\cite{Farajollahi3}, \cite{Nojiri}-\cite{Lopez}.

One way to generalize the gravitational action of a 5D theory is to bring in an induced gravity correction through considering a 4D scalar curvature term in the brane action in addition to the matter Lagrangian in it. A well-known example of induced-brane gravity model is the DGP model \cite{Dvali}. In this model, our 4D brane is embedded into an infinite 5D Minkowskian bulk. Also, in DGP model the cosmological constant of the bulk and of the brane and the brane tension set to zero, simultaneously. DGP model consists of two separate branches depending on how the brane embeds into the bulk. These branches where distinguish with a parameter $\epsilon=\pm1$, have distinct characteristics. For instance, the case $\epsilon=+1$, called self-accelerating branch, induces a late-time acceleration without need to a dark energy component and the case $\epsilon=-1$, called normal branch against the prior needs a dark energy component to explain the late-time acceleration.

DGP model has been frequently utilized in studying the dynamics of the Universe in its dark dominated stages \cite{Yin}-\cite{Farajollahi8} and in the inflationary era \cite{Cai}-\cite{Herrera3}. In \cite{Herrera2} and \cite{Herrera3}, intermediate and warm intermediate inflation in the context of DGP cosmology but in the absence of tachyon field has been studied, respectively. On the other hand, tachyon as an inflaton field has been used in \cite{Farajollahi3}, in the context of intermediate inflation in brane cosmology, but not a DGP-brane scenario. In this manuscript we will investigate a tachyonic intermediate inflationary model in DGP cosmology. Our motivation to considering such a model in addition to fill the gap mentioned above is that all the pillars of our model, i.e., intermediate inflation, tachyon scalar field and the higher dimensional DGP model, come from string theory and may lead to new and interesting results.

The outline of this work is as follows. In the next section we start with the action of the DGP model. Considering the slow-roll inflationary conditions and in the high energy regime the effective Friedmann equation and KG equation of the tachyon field will be obtained. After introducing the slow-roll parameters, in section III, we derive some important parameters related to perturbation theory in our model. Section IV, deals with numeric approaches to test the validity of our model. To this aim we use recent observational constraints from Planck satellite. The last section will demonstrate a summary of our work and its results.

\section{The model}

Our starting point is the action of DGP brane-world model which can be written as
\begin{equation}\label{action}
S=\frac{1}{2\kappa^2_{5}}\int d^5x\sqrt{-g_5}{R_5}+\int d^4x\sqrt{-g_4}{\cal{L}} .
\end{equation}
The first term in the above is corresponding to the Einstein-Hilbert action in a 5D Minkowskian bulk and the second term is the contribution of induced gravity localized on the brane. Here, $R_5$ is the 5D Ricci scalar and ${\cal{L}}$ is the effective 4D Lagrangian on the brane which can be expressed as
\begin{equation}\label{lb}
{\cal{L}}=\frac{\mu^2}{16\pi}R_4+L_{m},
\end{equation}
where $\mu$ is a mass parameter controlling the strength of the induced gravity term, which may correspond to the 4D Planck mass, $M_4$. Also, $R_4$ and $L_m$ are the Ricci scalar and the matter Lagrangian on the brane, respectively. In fact, our action is a special case of a more general induced gravity action in which the cosmological constants in the bulk and on the brane and the tension of the brane have been set to zero.

Considering the most relevant spatially flat FRW metric on the brane and introducing $\rho_0=\frac{48\pi}{\kappa_{5}^4\mu^2}$, we obtain the Friedmann equation of our model as
\begin{equation}\label{fried}
H^2=\frac{8\pi}{3\mu^2}(\sqrt{\rho+\frac{\rho_0}{2}}+\epsilon\sqrt{\frac{\rho_0}{2}})^2,
\end{equation}
in which $\epsilon$ can take the values $\pm$1, as we mentioned earlier. Since inflation is a period in the very early universe, we impose the high energy condition $\rho\gg\rho_0$ \cite{Maeda};\cite{Lopez3} in our model. For latter convenience we rewrite the effective Friedmann equation in the inflationary era as
\begin{equation}\label{he}
H^2=\frac{8\pi}{3\mu^2}(\sqrt\rho+\epsilon\sqrt{\frac{\rho_0}{2}})^2.
\end{equation}
One can check that equation (\ref{he}) is a suitable approximation of the related effective Friedmann equation in literature \cite{Herrera2};\cite{Maeda}.

Apparently, from the action, the matter is only confined to the brane, so it obeys the standard form of conservation equation
\begin{equation}\label{cons}
 \dot{\rho}+3H(\rho+p)=0.
\end{equation}
In the following we will consider a tachyon scalar field as the matter on the brane which plays the role of inflaton field in the inflationary era. For a tachyon field, the energy density and the pressure are given by
\begin{eqnarray}\label{rho}
\rho_\phi=\frac{V(\phi)}{\sqrt{1-\dot\phi^2}}, \quad
p_\phi=-V(\phi)\sqrt{1-\dot{ \phi^2}},
\end{eqnarray}
where $V(\phi)$, is the tachyon  potential. Replacing Eqn.(\ref{rho}) in (\ref{cons}), one obtains the equation of motion of the tachyon field as,
\begin{equation}\label{field}
\frac{\ddot {\phi}}{1-\dot\phi^2}+3H\dot \phi +\frac{V'(\phi)}{V}=0,
\end{equation}
in which $V'=\partial V(\phi)/\partial \phi$. Using equations (\ref{he}), (\ref{cons}) and (\ref{rho}), we reach to
\begin{equation}\label{dphi}
\dot\phi^2=\frac{-2\dot H}{3H^2\left(1-\frac{\alpha}{H}\right)}\cdot
\end{equation}
in which $\alpha=\epsilon\sqrt{\frac{4\pi\rho_0}{3\mu^2}}$. Considering intermediate inflationary scenario with the scale factor $a(t)=\exp{(At^f)}$, we obtain a solution for the above equation as
\begin{equation}\label{phi}
\phi(t)-\phi_0=C{\cal{B}}(t),
\end{equation}
where $\phi_0$ is the integration constant and means $\phi(t=0)$. Also, $C$ is another constant as below
\begin{equation}\label{c}
    C=\left[\frac{2}{3Af(1-f)}\beta^{\frac{f-2}{1-f}}\right]^{1/2},
\end{equation}
in which $\beta=\alpha/(Af)$ and ${\cal{B}}(t)$ represents the following incomplete Beta function \cite{Arfken}
\begin{equation}\label{beta}
    {\cal{B}}(t)=B\left[\beta t^{1-f};\frac{2-f}{2(1-f)},\frac{1}{2}\right].
\end{equation}

Without loss of generality, we assume $\phi_0=0$, and by using Eqn.(\ref{phi}), we find the Hubble parameter as a function of tachyon field
\begin{equation}\label{Hphi}
    H(\phi)=Af\left[{\cal{B}}^{-1}\left(\frac{\phi}{C}\right)\right]^{f-1},
\end{equation}
where ${\cal{B}}^{-1}$ represents the inverse function of the incomplete Beta function.

Also, using Eqns.(\ref{he}), (\ref{rho}) and (\ref{dphi}), the effective potential of our model can be obtained as
\begin{equation}\label{v}
V=\frac{3\mu^2H^2}{8\pi}\left(1-\frac{\alpha}{H}\right)^2\left[1+\frac{2\dot H}{3H^2(1-\frac{\alpha}{H})}\right]^{1/2}.
\end{equation}

Supporting a long enough period of inflation, the tachyon field must slowly rolls down its potential. In this scenario which is called slow-roll inflation the energy density of the inflaton field and its potential satisfy $\rho_{\phi}\sim V$. If the inflaton field is a tachyon field as in our model, the slow-roll conditions will be $\dot \phi^2\ll 1$ and $\ddot \phi\ll 3H \dot \phi$. Under these conditions, Eqns.(\ref{he}) and (\ref{field}) reduce to
\begin{equation}\label{fried2}
\frac{3\mu^2H^2}{8\pi}\left(1-\frac{\alpha}{H}\right)^2\approx V,
\end{equation}
and
\begin{eqnarray}\label{field2}
\frac{V'}{V}\approx -3H\dot \phi,
\end{eqnarray}
respectively. Also, the effective tachyon potential, Eqn.(\ref{v}), as a function of tachyon scalar field, becomes
\begin{equation}\label{slowv}
    V(\phi)=\frac{3\mu^2A^2f^2}{8\pi}\left[\left({\cal{B}}^{-1}(\frac{\phi}{C})\right)^{f-1}-\beta\right]^2.
\end{equation}

Slow-roll parameters are a few dimensionless parameters one can introduce in any slow-roll inflationary model. In our model, they will be as
\begin{equation}\label{epsilon}
\varepsilon=-\frac{\dot H}{H^2}= \frac{(1-f)}{Af}\left[{\cal{B}}^{-1}\left(\frac{\phi}{C}\right)\right]^{-f},
\end{equation}
and
\begin{equation}\label{eta}
\eta=-\frac{\ddot H}{H\dot H}= \frac{(2-f)}{Af}\left[{\cal{B}}^{-1}\left(\frac{\phi}{C}\right)\right]^{-f}\cdot
\end{equation}
The inflationary phase takes place whenever $\ddot a> 0$, which is proportional to $\varepsilon < 1$. Consequently, in our model, $\phi>C{\cal{B}}[(\frac{1-f}{Af})^{1/f}]$ is the necessary and sufficient condition for inflation to occur. Also, if we consider that the inflation begins at the earliest possible phase, $(t=t_1)$, in which $\varepsilon =1$ \cite{Barrow};\cite{Herrera4}, we obtain, $\phi(t=t_1)=\phi_1=C{\cal{B}}[(\frac{1-f}{Af})^{1/f}]$.

The number of $e$-folds between two cosmological times $t_1$ and $t_2>t_1$ is defined as the logarithm of the amount of expansion between them. In our model, it can be expressed as
\begin{equation}\label{N}
N=\int_{t_1}^{t_2} Hdt= A(t_2^f-t_1^f)=A\left[\left({\cal{B}}^{-1}\left(\frac{\phi_2}{C}\right)\right)^f-\left({\cal{B}}^{-1}\left(\frac{\phi_1}{C}\right)\right)^f\right].
\end{equation}

\section{Perturbation}

 While homogeneous and isotropic Universe assumption is valid in studying cosmology, in recent observations the existence of some deviations from this assumption are undoubted. Therefore, it seems necessary to consider the perturbation analysis of the model . The attractive feature of gravity can cause growing the inhomogeneities with time and since they are very small in early universe one can assume a linear perturbation scenario in inflation period. Here, we focus on a perturbed inflaton field in a perturbed geometry.

The most general linearly perturbed flat FRW metric which includes both scalar and tensor perturbations can be written as
\begin{equation}\label{metric}
ds^2 = -(1+2C)dt^2+2a(t)D_{,i}dx^idt+a(t)^2[(1-2\psi)\delta_{ij}+2E_{,i,j}+2h_{ij}]dx^idx^j,
\end{equation}
where $C, D, \psi $ and $E$ are the scalar metric perturbations and $h_{ij}$ is the transverse-traceless
tensor perturbation. For describing the distinctive nature of perturbations, we use both power spectrum of the curvature and tensor perturbation, ${\cal P}_{\cal R}$ and ${\cal P}_g$. These quantities appear in deriving the correlation function of the inflaton field in the vacuum state.

For the tachyon field, ${\cal P}_{\cal R}$, is defined as ${\cal P}_{\cal R}=(\frac{H^2}{2\pi\dot\phi})^2\frac{1}{Z_s}$, where $Z_s=V(1-\dot\phi^2)^{-3/2}$ \cite{Hwang}. In slow-roll approximation, it reduces to ${\cal P}_{\cal R}\approx(\frac{H^2}{2\pi\dot\phi})^2\frac{1}{V}$ and by using Eqns.(\ref{dphi}) and (\ref{fried2}), it becomes
\begin{equation}\label{pr}
{\cal P}_{\cal R}\approx\frac{-H^4}{\pi\mu^2\dot H\left(1-\frac{\alpha}{H}\right)}\cdot
\end{equation}
Now, considering intermediate inflation and with attention to Eqn.(\ref{phi}), the above relation can be rewritten in terms of tachyon field
\begin{equation}\label{pr1}
{\cal P}_{\cal R}\approx\frac{A^3f^3\left[{\cal{B}}^{-1}\left(\frac{\phi}{C}\right)\right]^{3f-2}}{\pi\mu^2(1-f)\left(1-\beta\left[{\cal{B}}^{-1}\left(\frac{\phi}{C}\right)\right]^{1-f}\right)}
\end{equation}
and consequently in terms of the number of $e$-folds, $N$, as below
\begin{equation}\label{pr1}
    {\cal P}_{\cal R}\approx\frac{A^3f^3\left[\frac{N}{A}+\frac{1-f}{Af}\right]^\frac{3f-2}{f}}{\pi\mu^2(1-f)\left(1-\beta\left[\frac{N}{A}+\frac{1-f}{Af}\right]^\frac{1-f}{f}\right)},
\end{equation}
where we have used Eqn.(\ref{N}). Similar to \cite{Herrera4} and \cite{Lopez2}, we can introduce a new parameter $\gamma=\mu^2/M_4^2$, where $0\leq\gamma<1$, is a dimensionless constant. In a model without induced gravity correction such as RS model, we have $\gamma=0$. Using Eqn.(\ref{pr1}) and for given values of $\epsilon$, $f$, $\gamma$, $M_4$, $\rho_0$, $N$ and ${\cal P}_{\cal R}$, we can find a constraint on the parameter $A$, numerically. In the following we will work in the normal branch of DGP model in which $\epsilon=-1$. Later, we will explain why we have neglected the case $\epsilon=+1$. Also, in \cite{Cai}, applying low energy condition into the Friedmann equation, the authors have shown that the parameter $\rho_0$, should be at least in the order of $(10^{-3} eV)^4$, as the current critical energy density of our Universe. Using ${\cal P}_{\cal R} \simeq 2.4\times10^{-9}$ and $N\simeq60$ and assuming $M_4=1$ and $\gamma=0.5$, we found constraints on $A$, related to different values of $f$. The results have been indicated in table \ref{table:1}.

\begin{table}[h]
\caption{Constraints on $A$} 
\centering 
\begin{tabular}{|ccc|c|c|c|c|} 
\hline 
\hline
 & $f$ &  &  0.1 & 0.5 & 0.7 & 0.9 \\ [1ex] 
\hline
 & $A$ & & 19.478 & $3.097\times10^{-2}$ & $8.756\times10^{-4}$ & $1.579\times10^{-5}$ \\ [1ex]
\hline
\end{tabular}
\label{table:1} 
\end{table}

On the other hand, tensor perturbation would produce gravitational
waves during inflation. Since an extra dimension allows gravitons to propagate into the bulk, the tensor perturbation is more important in our model. The amplitude of tensor perturbations in an induced gravity model has been calculated in \cite{Lopez2} and \cite{Lidsey},
\begin{equation}\label{power-tensorper}
{\cal P}_g=\frac{64\pi}{M_4^2}\left(\frac{H}{2\pi}\right)^2 G_\gamma^{2}(x).
\end{equation}
This form of amplitude of tensor perturbations differs from its expression in a standard 4D general theory of relativity by the coefficient $G_\gamma^{-2}(x)=\gamma+(1-\gamma)F(x)^{-2}$, Here, $F(x)$ is defined as
\begin{equation}\label{f}
F(x)=\left[\sqrt{1+x^2}-x^2\sinh^{-1}(1/x)\right]^{-1/2},
\end{equation}
where $x=H/\bar\mu$, and $\bar\mu$ is the energy scale associated with the bulk curvature. Noting that, for derving the amplitude of tensor perturbations one needs the amplitude of the zero-mode metric fluctuations on the brane. From \cite{Lopez2}, in a brane-world inflation with induced gravity, the zero-mode is not normalizable in the positive branch, $\epsilon=+1$. As a result, in here we restrict our calculations to the negative or normal branch, $\epsilon=-1$.

 Since we are working with a original DGP model in which the bulk is Minkowskian, we have $\bar\mu\rightarrow0$. In another word, the energy scale at which inflation begins will satisfies the condition $H\gg\bar\mu$. According to \cite{Lopez2}, under this condition, Eqn.(\ref{power-tensorper}), reduces to
\begin{equation}\label{power-tensorper2}
{\cal P}_g\approx\frac{64\pi}{M_4^2}\left(\frac{H}{2\pi}\right)^2\frac{1}{\gamma}.
\end{equation}
Therefore, in the context of intermediate inflation we obtain
\begin{equation}\label{power-tensorper3}
    {\cal P}_g\approx2\left(\frac{Af}{\pi\mu}\right)^2\left[{{\cal{B}}^{-1}\left(\frac{\phi}{C}\right)}\right]^{2(f-1)}.
\end{equation}
Another useful quantity in studying perturbation theory is tensor-to-scalar ratio
\begin{equation}\label{r}
r=\frac{{\cal P}_g}{{\cal P}_{\cal R}},
\end{equation}
which is the ratio between power spectrum of tensor perturbation and power spectrum of scalar perturbation. In our model Eqn.(\ref{r}), reduces to
\begin{equation}\label{r2}
    r\approx\frac{16(1-f)}{ Af}\left(1-\beta\left[{{\cal{B}}^{-1}\left(\frac{\phi}{C}\right)}\right]^{1-f}\right)\left[{{\cal{B}}^{-1}\left(\frac{\phi}{C}\right)}\right]^{-f}\cdot
\end{equation}

Identification of two other parameters in the topic of scalar perturbation in cosmology is of particular interest. These are the scalar spectral index $n_s$, which is related to ${\cal P}_{\cal R}$, through the relation $n_s-1=d\ln {\cal P}_{\cal R}/d\ln k$, and the running in the scalar spectral index parameter $n_{run}$, which shows the scale dependence of primordial scalar fluctuations and can be obtained via $n_{run}=d n_s/d \ln k$. Here, $d\ln k=dN$.

In our model and under slow-roll approximation, they, respectively, become
\begin{equation}\label{ns}
n_s\approx1-\frac{2-3f}{Af}\left[{{\cal{B}}^{-1}\left(\frac{\phi}{C}\right)}\right]^{-f}+\frac{\beta(1-f)}{Af\left(1-\beta\left[{{\cal{B}}^{-1}\left(\frac{\phi}{C}\right)}\right]^{1-f}\right)}\left[{{\cal{B}}^{-1}\left(\frac{\phi}{C}\right)}\right]^{1-2f}
\end{equation}
and
\begin{eqnarray}\label{nrun}
n_{run}&\approx&\frac{2-3f}{A^2f}\left[{{\cal{B}}^{-1}\left(\frac{\phi}{C}\right)}\right]^{-2f}+\frac{\beta (1-f)(1-2f)}{A^2f^2}\frac{\left[{{\cal{B}}^{-1}\left(\frac{\phi}{C}\right)}\right]^{1-3f}}{\left(1-\beta\left[{{\cal{B}}^{-1}\left(\frac{\phi}{C}\right)}\right]^{1-f}\right)}\\ \nonumber
&+& \frac{\beta^2(1-f)^2}{A^2f^2}\frac{\left[{{\cal{B}}^{-1}\left(\frac{\phi}{C}\right)}\right]^{2-4f}}{\left(1-\beta\left[{{\cal{B}}^{-1}\left(\frac{\phi}{C}\right)}\right]^{1-f}\right)^2}\cdot
\end{eqnarray}

Noting that, from equation (\ref{ns}), it is clear that the Harrison-Zel'dovich model, i.e., $n_s=1$, cannot be obtained for $f=2/3$, as occurs in a standard tachyonic intermediate inflation \cite{Campo2}. Also, unlike the situation in a tachyonic brane intermediate inflationary model \cite{Farajollahi3}, we do not achieve $n_s=1$, for $f=3/4$.

\section{Numerical discussion}

In this section we do some numeric calculations to check the consistency of our model. To this aim, we plot some trajectories in $(r - n_s)$ and $(n_{run} - n_s)$ planes obtained from our model and compare them with confidence regions extracted from \cite{Planck1};\cite{Planck2}. We know that in the standard cosmology we apply a spatially-flat six-parameter $\Lambda$CDM model. These parameters include baryon density today ($\Omega_b h^2$), cold dark matter density today ($\Omega_c h^2$), angular scale of the sound horizon at last-scattering ($\theta_{s}$), Thomson scattering optical depth due to reionization ($\tau$), scalar spectrum power-law index ($n_s$) and log power of the primordial curvature perturbations ($\ln(10^{10}A_s)$).

In \cite{Planck2}, using the combination of Planck temperature (Planck TT) and Planck polarization (lowP) data, in $\Lambda$CDM model, the authors have found at the 68\% confidence level $n_s=0.9655\pm0.0062$, at the pivot scale $k_*=0.05$Mpc$^{-1}$, which is one of the largest shifts among these six parameters in comparison to Planck 2013 results. Several improvements in the data processing shifts $n_s$, towards higher values. For instance, adding baryon acoustic oscillations (BAO) to Planck TT+lowP, causes $n_s=0.9673\pm0.0045$. Also, $n_s=0.9677\pm0.0060$ in the case Planck TT+lowP+lensing.

Although the $\Lambda$CDM model is in good agreement with observations, but a simple one-parameter extension of it, called $\Lambda$CDM+$r$ model, which considers the contribution of tensor perturbations is of particular interest. According to inflationary theories the primordial tensor fluctuations or gravitational waves contribute to temperature and polarization anisotropy of CMB. It has been shown in \cite{Planck2} that, considering Planck TT+lowP, there is a movement to slightly higher values for $n_s$, in $\Lambda$CDM+$r$ model. Also it has been shown in \cite{Planck1} that adding Planck lensing, BAO and some other astrophysical data such as the joint light-curve analysis (JLA) sample of Type Ia supernovae and Hubble constant direct measurement data from Hubble Space Telescope ($H_0$), to Planck TT+lowP in $\Lambda$CDM+$r$ model, leads to tighter constraints in $n_s$ direction in $(r-n_s)$ plane and again shift of $n_s$, to higher values \cite{Planck2}. It also causes an increase in the 95\% upper limit of $r_{0.002}$, from 0.10 to 0.11 \cite{Planck2}, though adding the joint analysis of BICEP2, Keck Array and Planck polarization data (BKP) \cite{BKP} to this collection decreases it to 0.09 \cite{Planck1}. In here, the subscript 0.002 is related to $k_*=0.002$Mpc$^{-1}$.

From CMB experiments, with and without combination of some other astrophysical data, some hints of a non-zero running with a slight preference for negative values have been found \cite{Planck2}. Considering the contribution of both gravitational waves and running in a model dubbed $\Lambda$CDM+$r$+$n_{run}$, leads to $n_{run}=-0.0126^{+0.0098}_{-0.0087}$ using Planck TT+lowP \cite{Planck2}, and $n_{run}=-0.0065\pm0.0076$ using Planck TT+lowP+lensing+BAO+JLA+$H_0$+BKP \cite{Planck1}. These values have been resulted at $k_*=0.05$Mpc$^{-1}$. Adding a running of the scalar spectral index as an additional free parameter in $\Lambda$CDM+$r$ model weakens the upper limit on $r$, where again this limit can be decreased if we consider the BKP data, as well \cite{Planck1}. Also, running causes a movement of $n_s$, towards higher values in some cases \cite{Planck2}.

The $\Lambda$CDM+$r$+$n_{run}$ model significantly reduced a tension in recent past about determining the constraint on the tensor-to-scalar ratio parameter, $r$, between the BICEP2 and PT13 \cite{Cheng}. However, a number of studies such as the joint analysis of BKP, has removed this tension without need of considering the scalar spectral index running \cite{BKP}.

On the other hand, some parameters related to neutrino physics can be used to make an extension of the $\Lambda$CDM model \cite{Planck1}. In one of these extended models, two new parameters, the effective number of massive and massless neutrinos $N_{\textrm{eff}}$, and the effective sterile neutrino mass $m_{\nu,\textrm{sterile}}^{\textrm{eff}}$, are added simultaneously to the $\Lambda$CDM+$r$ model. We represent this model with $\Lambda$CDM+$r$+$\nu_s$, as it has been shown in \cite{Zhang}. In addition to reducing the tension about $r$, between PT13 and BICEP2 results, this model can strongly relax the other tensions between Planck and some of the local astrophysical data, such as $H_0$, Sunyaev-Zeldovich cluster counts data and the Cosmic Shear data \cite{Zhang}. The cost of reducing these tensions in $\Lambda$CDM+$r$+$\nu_s$ model is an increasing in the value of $n_s$ \cite{Zhang}. In the following we compare our model with these two extensions of the $\Lambda$CDM model, separately.

\subsection{$\Lambda$CDM+$r$+$\nu_s$ model}

\begin{figure}[t]
\centering
\includegraphics[width=0.45\textwidth]{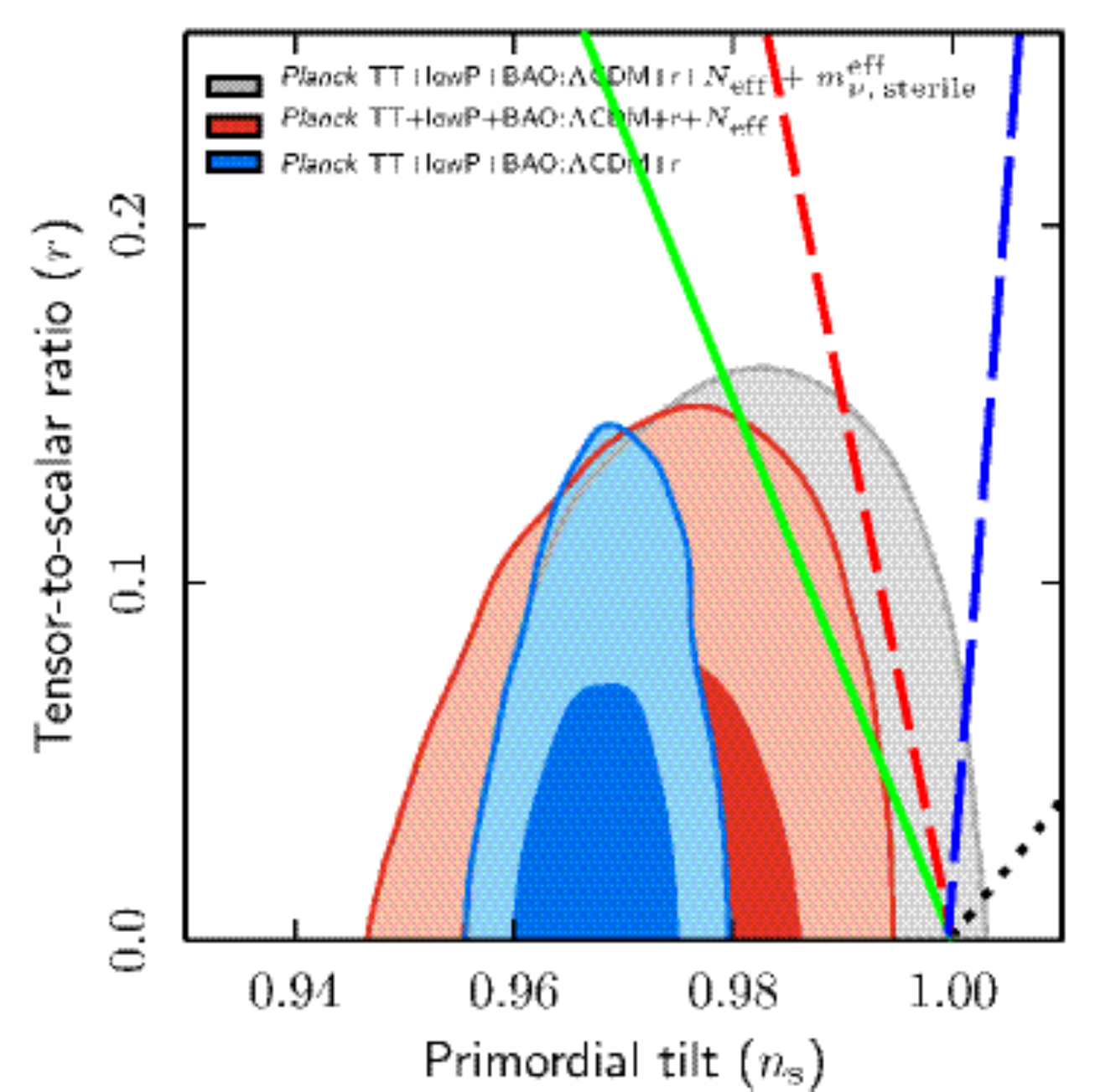}
\caption{The comparison between the curves of r($n_s$) in our model and observational data. The solid green, dashed red, long dashed blue and dotted black curves, have obtained from our model and are related to $f = 0.1, 0.5, 0.7$ and $0.9$, respectively. The colored contours show 68\% and 95\% confidence regions obtained from Planck TT+lowP+BAO data sets, in a $\Lambda$CDM+$r$ (blue), $\Lambda$CDM+$r$+$N_{eff}$ (red) and $\Lambda$CDM+$r$+$\nu_s$ (gray) model. Our model fits the observation well for $f<0.7$, if we consider the $\Lambda$CDM+$r$+$\nu_s$ model.}\label{fig2}
\end{figure}

First of all, we address to $\Lambda$CDM+$r$+$\nu_s$ model in which three new parameters, $r$, $N_{\textrm{eff}}$ and $m_{\nu,\textrm{sterile}}^{\textrm{eff}}$, have been added to the standard $\Lambda$CDM model. This model can significantly reduce all the tensions between Planck and some other observational data.
In Fig.(\ref{fig2}), we have compared the trajectories in $(r - n_s)$ plane of our model with the related confidence regions from \cite{Planck2}, in which the authors have used Planck TT+lowP+BAO data to perform their numerical analysis. In this figure the blue, red and gray contours are related to the case of considering $\Lambda$CDM+$r$, $\Lambda$CDM+$r$+$N_{eff}$ and $\Lambda$CDM+$r$+$\nu_s$, respectively. It is obvious from this figure that, in the case of the $\Lambda$CDM+$r$+$\nu_s$ model for $f<0.7$, our model fits the observations well, .

\subsection{$\Lambda$CDM+$r$+$n_{run}$ model}

\begin{figure}[t]
\centering
\includegraphics[width=0.45\textwidth]{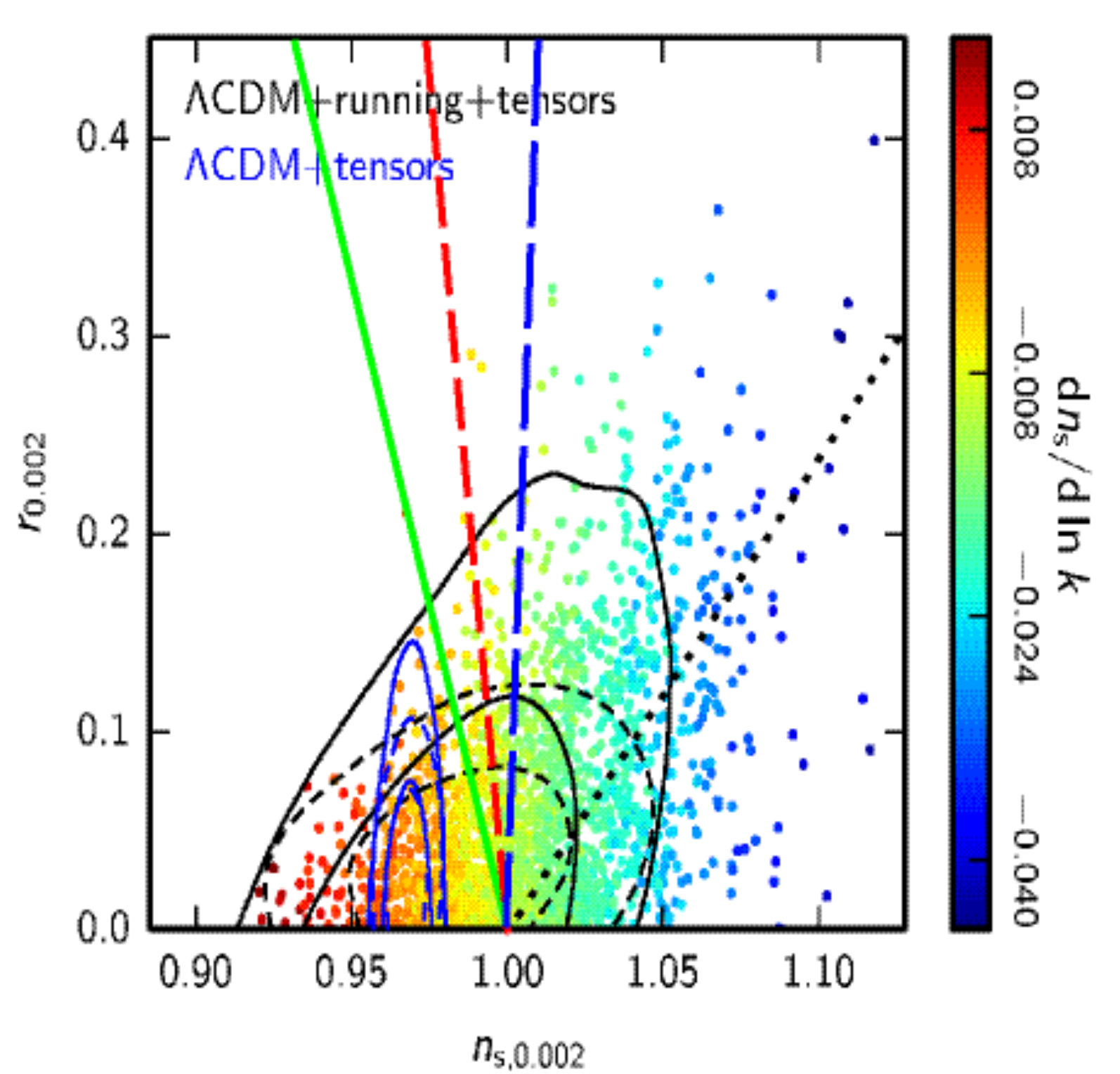}
\caption{The comparison between the curves of r($n_s$) in our model and observational data. The solid green, dashed red, long dashed blue and dotted black curves, have obtained from our model and are related to $f = 0.1, 0.5, 0.7$ and $0.9$, respectively. The blue and black contours show 68\% and 95\% confidence regions in a $\Lambda$CDM+$r$ and a $\Lambda$CDM+$r$+$n_{run}$ model, respectively. Also, the solid contours obtained using Planck TT+lowP+lensing+BAO data sets and the dashed contours are related to the case of adding BKP data to priors. If we only use Planck TT+lowP data, we reach to the colored samples. Our model fits the observation very well in a $\Lambda$CDM+$r$+$n_{run}$ model.}\label{fig3}
\end{figure}

Next, we address the $\Lambda$CDM+$r$+$n_{run}$ model in which in addition to $r$, the new parameter $n_{run}$ has been added to the $\Lambda$CDM model. As we mentioned earlier this model can significantly relieve some tensions between astrophysical data \cite{Cheng}. We are interested in this model, because in addition to $(r - n_s)$ plane, we can do one more test of our model using $(n_{run} - n_s)$ plane. In Fig.(\ref{fig3}), the solid contours have been illustrated using the Planck TT+lowP+lensing+BAO data. Comparing our results with these confidence regions obtained in \cite{Planck1}, we find that our model works very well if we consider the effect of running, even if we add the BKP data to prior datasets (the dashed contours).

Also, the colored points in Fig.(\ref{fig3}) are samples that colored by running parameter and are related to the case when we only consider Planck TT+lowP data in $\Lambda$CDM+$r$+$n_{run}$ model. It is clear from these samples that an increase in $n_s$ parameter is equivalent to a decrease in $n_{run}$.

\begin{figure}[t]
\centering
\includegraphics[width=0.45\textwidth]{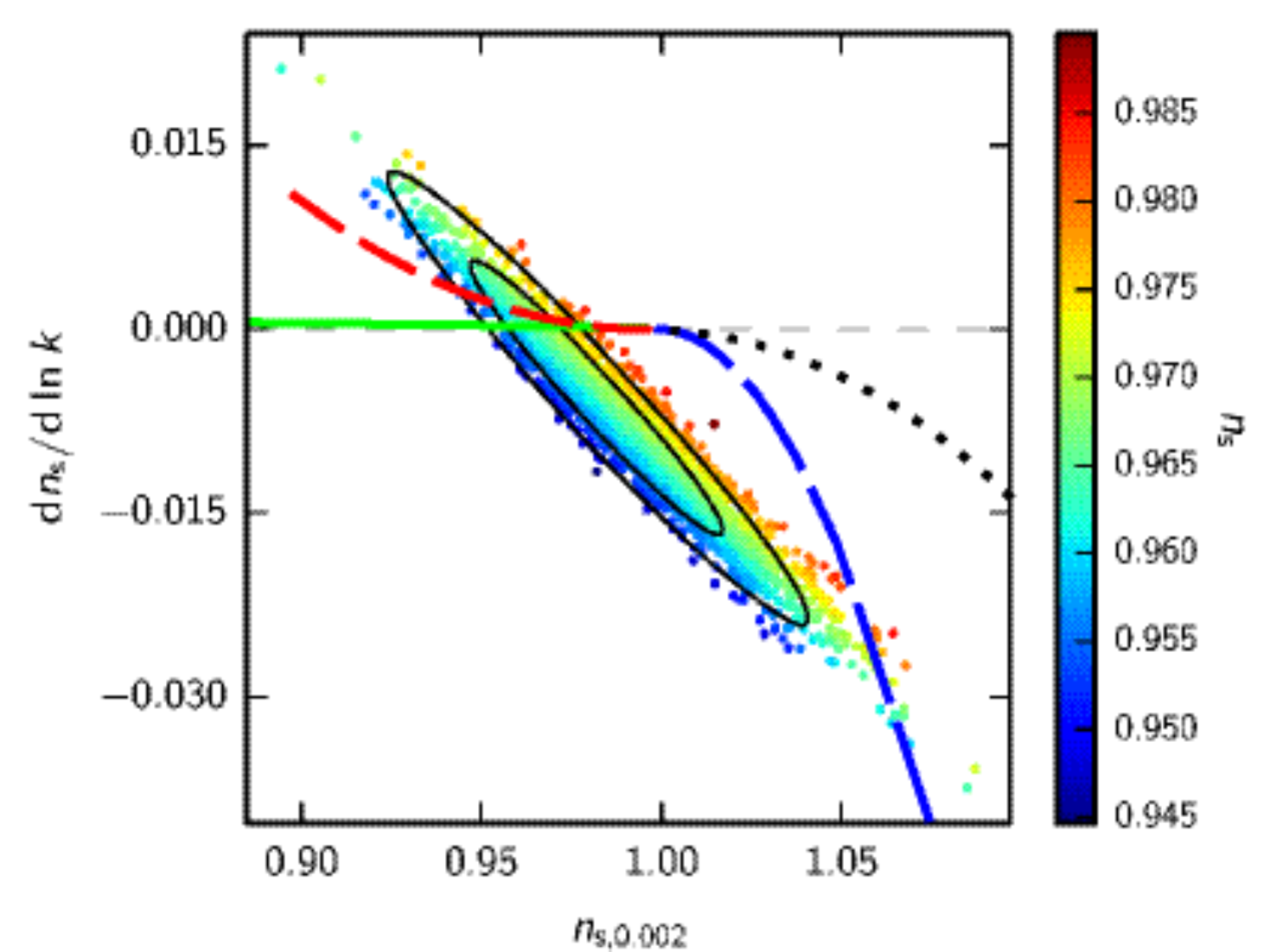}
\caption{The comparison between the curves of $n_{run}$($n_s$) in our model and observational data. The solid green, dashed red, long dashed blue and dotted black curves, have obtained from our model and are related to $f = 0.1, 0.5, 0.7$ and $0.9$, respectively. The black contours show 68\% and 95\% confidence regions in a $\Lambda$CDM+$r$+$n_{run}$ model, obtained from Planck TT,TE,EE+lowP data sets. If we only use Planck TT+lowP data at $k_*=0.05$Mpc$^{-1}$, we reach to the colored samples. Our model fits the observation well for $f<0.7$.}\label{fig4}
\end{figure}

Though in Fig.(\ref{fig3}), we see that our model works well for any value of $f$, but, with attention to $(n_{run} - n_s)$ plane in Fig.(\ref{fig4}), we deduce that our model is consistent with observations for the values $f<0.7$. Then, we conclude that in the case of a $\Lambda$CDM+$r$+$n_{run}$ model, our tachyonic intermediate inflationary scenario fits observations for $f<0.7$. In Fig.(\ref{fig4}), the colored samples is related to $n_s$ at $k_*=0.05$Mpc$^{-1}$, obtained using Planck TT+lowP data, but, the black contours come from Planck TT,TE,EE+lowP.

\section{Summary}

In this work we have analyzed a tachyonic intermediate inflationary model in the context of the DGP cosmology. The main motivation for considering this model is that all the pillars of our work, i.e., an intermediate inflationary model, a tachyon scalar field and the DGP cosmology, come from string theory. Applying slow-roll conditions and in high energy regime we obtained the effective Friedmann and KG equations in our model. Considering a general perturbed FRW metric, we derived the explicit expressions of some parameters in perturbation theory, such as power spectrum of scalar perturbation, ${\cal{P}}_{\cal{R}}$, power spectrum of tensor perturbation, ${\cal{P}}_g$ and the ratio between them, $r$. Also, we obtained the scalar spectral index, $n_s$ and its running $n_{run}$, in terms of the tachyon field.

Then we did some numeric calculations to compare our model with observations. We illustrated the trajectories $r - n_s$ and $n_{run} - n_s$ in our model by using the related confidence regions from \cite{Planck1} and \cite{Planck2}. Different extensions of the $\Lambda$CDM cosmological model have been investigated by utilizing distinct observational data sets to perform analysis related to perturbations. The important feature of these extended models is that they can relieve the tensions between some of the observational data. From Fig.(\ref{fig2}), it is obvious that in $\Lambda$CDM+$r$+$\nu_s$ model, our model fits the observations well for $f<0.7$. Although in the case of considering running of the scalar spectral index, from Figs. (\ref{fig3}) and (\ref{fig4}) simultaneously, we obtain the same result, i.e., the model is consistent with observations for $f<0.7$. In all these cases, for $f<0.7$ we have $n_s<1$. It seems that the closest scenario in our model to the Harrison-Zel'dovich model with $n_s=1$, is reached for $f=0.6$.

\end{document}